\documentclass{aa}

\usepackage{graphicx}
\usepackage{xcolor}
\usepackage{txfonts}
\usepackage[colorlinks=false]{hyperref}
\newcommand{\munu}{{\mu \nu}}

\newcommand{\Mpch}{\,{\rm Mpc}\,\ifmmode h^{-1}\else $h^{-1}$\fi}
\newcommand{\Msh}{\,\ifmmode M_\odot\,h^{-1}\else $M_\odot\,h^{-1}$\fi}

\begin{document} 

   \title{Breaking cosmic degeneracies: Disentangling neutrinos and modified gravity with kinematic information}
   \titlerunning{Breaking cosmic degeneracies with kinematics}

\author{
Steffen Hagstotz$^{1,2,3}$\thanks{E-mail: steffen.hagstotz@fysik.su.se},
Max Gronke$^{4,5}$\thanks{Hubble fellow},
David F. Mota$^{5}$
and Marco Baldi$^{6,7,8}$
}

   \institute{
Oskar Klein Centre, Department of Physics, Stockholm University, SE-106 91 Stockholm, Sweden
\and
Excellence Cluster Universe, Boltzmannstr. 2, 85748 Garching, Germany
\and
Universit\"ats-Sternwarte, Fakult\"at f\"ur Physik, Ludwig-Maximilians Universit\"at M\"unchen, Scheinerstr. 1, 81679 M\"unchen, Germany
\and
Department of Physics, University of California, Santa Barbara, CA 93106, USA
\and
Institute of Theoretical Astrophysics, University of Oslo, Postboks 1029 Blindern, 0315 Oslo, Norway
\and
Dipartimento di Fisica e Astronomia,
  Alma Mater Studiorum Universit\`{a} di Bologna, via Gobetti 93/1,
  40129 Bologna, Italy
  \and
  Astrophysics and Space Science
  Observatory Bologna, via Gobetti 93/2, 40129, Bologna, Italy
  \and
INFN - Sezione di Bologna, viale Berti Pichat 6/2, 40127,
  Bologna, Italy}
             
\authorrunning{S. Hagstotz et al.}

   \date{Draft from \today}


  \abstract{
    Searches for modified gravity in the large-scale structure try to detect the enhanced amplitude of density fluctuations caused by the fifth force present in many of these theories. Neutrinos, on the other hand, suppress structure growth below their free-streaming length. Both effects take place on comparable scales, and uncertainty in the neutrino mass leads to a degeneracy with modified gravity parameters for probes that are measuring the amplitude of the matter power spectrum.
    We explore the possibility to break the degeneracy between modified gravity and neutrino effects in the growth of structures by considering kinematic information related to either the growth rate on large scales or the virial velocities inside of collapsed structures.
    In order to study the degeneracy up to fully non-linear scales, we employ a suite of $N$-body simulations including both $f(R)$ modified gravity and massive neutrinos.
    Our results indicate that velocity information provides an excellent tool to distinguish massive neutrinos from modified gravity. Models with different values of neutrino masses and modified gravity parameters possessing a comparable matter power spectrum at a given time have different growth rates. This leaves imprints in the velocity divergence, which is therefore better suited than the amplitude of density fluctuations to tell the models apart. In such models with a power spectrum comparable to $\Lambda$CDM today, the growth rate is strictly enhanced. We also find the velocity dispersion of virialised clusters to be well suited to constrain deviations from general relativity without being affected by the uncertainty in the sum of neutrino masses.
}  
  

   \maketitle

%

\section{Introduction}
Nearly two decades after the first measurements of the accelerated expansion of space \citep[e.g.][]{Riess_etal_1998,Perlmutter_etal_1999,Schmidt_etal_1998} the fact that about $70\%$ of the Universe's energy content is in a form with a negative equation of state of $w\approx -1$ has been confirmed in numerous measurements \citep{wmap9,Planck_2015_XIII}.
Nevertheless, the nature of this `dark energy' is as puzzling as it has been since its discovery. Tremendous efforts in modern cosmology go into determining the amount and possible time-evolution of this unknown component. It is particularly problematic that few well-motivated frameworks for its physical nature exist -- apart from a cosmological constant. Many ideas \citep[e.g.][]{Dvali_Gabadadze_Porrati_2000} have by now been ruled out or shown to be intrinsically unstable. While there are still theories around (and always will be, since the parameter space of many of them is very flexible), they appear more or less contrived.

It is also important to recall that gravity is the `odd' fundamental force, and a lot of implicit assumptions are being made when extrapolating our knowledge over several orders of magnitudes to vastly different conditions and scales. These two points are, in fact, the main motivations behind a class of modified gravity theories \citep{Amendola2010deto.book.....A,2012PhR...513....1C}. Since general relativity as a theory of gravity is unique under very general assumptions \citep{Lovelock72}, any modification introduces new physical degrees of freedom. These can lead to accelerated expansion, but also tend to enhance gravity on a perturbative level as so-called fifth forces. To pass observational bounds, any of these models have to involve a `screening mechanism' leading to negligible deviations in, e.g., the solar-system where the predictions of general relativity have been confirmed to high precision \citep[e.g.][]{Berotti2003Natur.425..374B, Will2006LRR.....9....3W}.

In this work, we will circumvent the discussion of what characterizes a scientific theory (as opposed to, for instance, an effective one), and will instead treat the screened modified gravity models considered as examples of a (much) larger group of models. They all possess the common property that in addition to the Newtonian gravitational force $F_{\mathrm{N}}$, another \textit{fifth force} component $F_{\mathrm{Fifth}}$ exists, which is suppressed by some screening mechanism in high-density (or high-curvature) environments.
This choice is motivated by the fact that screening occurs in a range of scalar- and vector-field theories with different physical reasons, and is in fact essentially required by a large class of theories in order not to violate local gravity measurements.
 Examples of screening mechanisms which are implemented in those theories include:
\begin{itemize}
\item \textit{Chameleon} \citep{Khoury_Weltman_2004} where the range of the fifth force is decreased in regions of high spacetime curvature, thus, effectively hiding the additional force,
\item \textit{Symmetron} \citep{Hinterbichler_Khoury_2010,Hinterbichler_etal_2011} in which the coupling of the scalar field is density dependent,
\item \textit{Vainshtein} screening \citep{1972PhLB...39..393V} where the screening effect is sourced by the second derivative of the field value, and
\item others such as screening through disformal coupling \citep{PhysRevD.48.3641}.
\end{itemize}

As already indicated above, a major problem in the search for a new theory of gravity is that $\Lambda$CDM gets so far only confirmed to higher and higher precision. While minor discrepancies between probes of the early and late Universe exist, especially in measurements of the Hubble parameter $H_0$ \citep[see e.g.][]{Riess_etal_2016, Planck_2015_XIII} and $\Omega_m$ or $\sigma_8$ \citep[e.g.][]{Hildebrandt_etal_2017}, no major tension between its predictions and the data has been found. Historically, however, we know that this does not mean that $\Lambda$CDM is correct but that either we have not yet found the right probe where tensions might arise, or we have to push the limits to higher precision. While the latter approach can well be fruitful (as shown by the high-precision measurements of, e.g., the perihelion precession of Mercury; \citealp{le1859lettre}) and is the preferred path taken by many next generation instruments such as EUCLID \citep{Euclid-y} and WFIRST \citep{WFIRST}, we will focus on the former path, and are thus interested in deviations on the $\gtrsim 10\%$ level.

Several observable signatures of screened modified gravity models have been suggested in the literature such as deviations in the halo mass function \citep{Schmidt_2010,Davis_etal_2012,Puchwein_Baldi_Springel_2013,Achitouv_etal_2016}, or the structure of the cosmic web \citep{2014JCAP...07..058F,2018A&A...619A.122H}. 
However one concern raised by several authors \citep[e.g.,][]{Motohashi:2012wc,He_2013,Baldi_etal_2014} is that massive neutrinos and beyond-$\Lambda$CDM models might be degenerate.

In this work, we want to investigate how kinematic information can be used to break these degeneracies. This paper is structured as follows: in Sec.~\ref{sec:method} we introduce  the screened modified gravity models studied, and briefly review the effect of neutrinos on structure formation. We will also describe our numerical simulations used to explore the joint effects numerically. In Sec.~\ref{sec:results} we present our results, before we conclude in Sec.~\ref{sec:conclucsion}.

\section{Method}
\label{sec:method}
This section briefly summarises the effects of modified gravity and massive neutrinos on the evolution of the density field. We also present the simulation suite used to study the combined effects in the fully non-linear regime.

\subsection{Review of modified gravity}
\label{sec:fR}

To work within a well-defined framework, in this paper we focus on $f(R)$ gravity. As a starting point we assume the generalised Einstein-Hilbert action\footnote{We adopt natural units $c = \hbar = 1$}
\begin{equation}
S = \int \mathrm d x^4 \sqrt{-g} \left( \frac{R + f(R)}{16 \pi G} + \mathcal{L}_m \right) \: ,
\end{equation}
where we introduced a function $f$ of the Ricci scalar $R$, the Lagrangian $\mathcal{L}_m$ contains all other matter fields and we recover standard general relativity (GR) if we choose the function to be a cosmological constant $f = - 2 \Lambda^\mathrm{GR}$. For this paper, we use instead the form established by \cite{Hu2007}
\begin{equation}
\label{eq:f_R}
f(R) = - 2 \Lambda \frac{R}{R + m^2} \: ,
\end{equation}
with a constant suggestively named $\Lambda$ and an additional scale $m^2$ that both have to be fixed later on. Assuming $m^2 \ll R$ lets us expand the function
\begin{equation}
\label{eq:f_R_approx}
f(R) \approx -2 \Lambda - f_{R0} \frac{\bar R_0^2}{R} \: ,
\end{equation}
with the background value of the Ricci scalar $\bar R_0$ today, and we defined the dimensionless parameter $f_{R0} \equiv - 2 \Lambda m^2 / \bar R_0^2$ that expresses the deviation from GR. We will return to the characteristic scale of $f_{R0}$ later, but typically $|f_{R0}| \ll 1$. The constant $\Lambda = \Lambda^\mathrm{GR}$ is then fixed to the measured value of the cosmological constant by the requirement to reproduce the standard $\Lambda$CDM expansion history established by observations. However, note that it no longer has the interpretation of a vacuum energy. The phenomenology of the theory in this limit is then set by $f_{R0}$ alone. This particular choice of parameters also implies that the background evolution is indistinguishable from a $\Lambda$CDM universe, but the growth of perturbations will differ.

To work out the perturbation equations, we vary the action with respect to the metric to arrive at the modified Einstein equations
\begin{equation}
\label{eq:modified_Einstein_eq}
G_\munu - f_R R_\munu - \left( \frac{f}{2} - \Box f_R \right) g_\munu - \nabla_\mu \nabla_\nu f_R = 8 \pi G T_\munu \: .
\end{equation}
The new dynamical scalar degree of freedom $f_R \equiv \mathrm d f / \mathrm d R$ is responsible for the modified dynamics of the theory. To obtain the equation of motion for this scalar field, we consider the trace of Eq.~\ref{eq:modified_Einstein_eq}
\begin{equation}
\label{eq:field_equation_fR}
\nabla^2 \delta f_R = \frac{a^2}{3} \Big( \delta R(f_R) - 8 \pi G \delta \rho_m \Big) \: ,
\end{equation}
where we assumed the field to vary slowly (the quasi-static approximation) and we consider small perturbations  $\delta f_R \equiv f_R - \bar f_R$, $\delta R \equiv R - \bar R$ and $\delta \rho_m \equiv \rho_m - \bar \rho_m$ on a homogeneous background. To get a Poisson-like equation for the scalar metric perturbation $2\psi = \delta g_{00} / g_{00}$ we take the time-time component of Eq.~\ref{eq:modified_Einstein_eq} to arrive at
\begin{equation}
\label{eq:Poisson_fR}
\nabla^2 \psi = \frac{16 \pi G}{3} a^2 \rho_m - \frac{a^2}{6} \delta R(f_R) \: ,
\end{equation}
that now also depends on the scalar field. Solving the non-linear Eqs.~\ref{eq:field_equation_fR} and \ref{eq:Poisson_fR} in their full generality requires $N$-body simulations, but it is interesting to consider two edge cases to get some insight into the phenomenology of the theory.

If the field is large, $|f_{R0}| \gg | \psi |$, we can expand
\begin{equation}
\delta R \simeq \left .\frac{\mathrm d R}{\mathrm d f_R} \right \rvert_{R = \bar R} \delta f_R \: ,
\end{equation}
and we can solve Eqs.~\ref{eq:field_equation_fR} and \ref{eq:Poisson_fR} in Fourier space to get
\begin{equation}
\label{eq:Poisson_large_field}
k^2 \psi(k) = -4 \pi G \left( \frac{4}{3} - \frac{1}{3} \frac{\mu^2 a^2}{k^2 + \mu^2 a^2} \right) a^2 \delta \rho_m(k) \: ,
\end{equation}
with the Compton wavelength of the scalar field $\mu^{-1} = (3 \mathrm d f_R / \mathrm d R)^{1/2}$. For $k \gg \mu$ the second term vanishes and we obtain a Poisson equation with an additional factor $4/3$. On the other hand, for $k \ll \mu$ we recover standard gravity. The Compton wavelength $\mu^{-1}$ therefore sets the interaction range of an additional fifth force that enhances gravity by $1/3$. This is the maximum possible force enhancement in $f(R)$, irrespective of the choice of the function in Eq.~\ref{eq:f_R}.

For field values $| f_{R0} | \ll | \psi | $, the two terms on the right hand side of Eq.~\ref{eq:field_equation_fR} approximately cancel, so we arrive at
\begin{equation}
\label{eq:delta_R_unscreened}
\delta R \approx 8 \pi G \delta \rho_m
\end{equation}
and we also recover the standard Poisson equation from Eq.~\ref{eq:Poisson_fR}. This is the \textit{Chameleon} screening mechanism mentioned above to restore GR in regions of high curvature.

We can get an estimate of the scale where this screening transition occurs by solving Eq.~\ref{eq:field_equation_fR} formally with the appropriate Green's function
\begin{align}
\label{eq:f_R_solution}
\delta f_R(r) &= \frac{1}{4 \pi r} \frac{1}{3} \int_0^r \mathrm d^3 \mathbf{r^\prime} 8 \pi G \left( \delta \rho - \frac{\delta R}{8 \pi G} \right) \\
 &= \frac{2}{3} \frac{G M_\mathrm{eff}(r)}{r}
\end{align}
where we defined the effective mass term $M_\mathrm{eff}$ acting as a source for the fluctuations in the scalar field $\delta f_R$. This definition requires $M_\mathrm{eff}(r) \leq M(r)$, and both contribution are equal in the unscreened regime, where Eq.~\ref{eq:delta_R_unscreened} implies $M_\mathrm{eff} = M$. In this case, $\delta f_R = 2/3 \psi_N$ with the Newtonian potential of the overdensity, $\psi_N = GM/r$. Since we assumed small perturbations on the homogeneous background, $\delta f_R \leq \bar{f_R}$, we arrive at the screening condition

\begin{equation}
\label{eq:thin_shell}
| f_{R}| \leq \frac{2}{3} \psi_N(r) \: .
\end{equation}

In other words, only the mass distribution outside of the radius where the equality $2/3 \psi(r) = |f_R|$ holds contributes to the fifth force. Note that screening for real halos is considerably more complex, since non-sphericity and environmental effects are also important for the transition. Nevertheless, Eq.~\ref{eq:thin_shell} gives a reasonable estimate for the onset of the transition between enhanced gravity and normal GR.

Since screening can function only for $\psi_N \sim f_R$, the condition implied by Eq.~\ref{eq:thin_shell} sets the scale for the free parameter $|f_{R0}|$. Typical values for the metric perturbation in cosmology range from $\psi_N \sim 10^{-5}$ to $\psi_N \sim 10^{-6}$, so $|f_{R0}|$ should be of the same order of magnitude to show any interesting phenomenology. For values of the scalar field $|f_{R0}| \gg \psi_N$, gravity is always enhanced so we can exclude this parameter space trivially, while in the opposite limit $|f_{R0}| \ll \psi$ the theory is always screened and does not offer any predictions to distinguish it from GR on cosmological scales.

\subsection{Neutrino effects on structure growth}

Cosmology allows to constrain the physics of neutrinos in unique ways. Assuming the standard thermal evolution and decoupling before $e^+/e^-$ annihilation, their temperature is related to the one of the CMB photons by
\begin{equation}
    T_\nu = \left( \frac{4}{11} \right)^{1/3} T_\mathrm{CMB} \: ,
\end{equation}
which implies for neutrinos with mass eigenstates $m_\nu$ a total contribution to the Universe's energy budget of \citep{MANGANO2005221}
\begin{equation}
\label{eq:Omega_nu}
\Omega_\nu h^2 \approx \frac{\sum m_\nu}{93.14~\mathrm{eV}} \: ,
\end{equation}
where the sum runs over the three standard model neutrino states. Since their mass is constrained to be small, $\sum m_\nu \lesssim 1~\mathrm{eV}$, they decouple as highly relativistic particles in the early Universe. Their energy density therefore scales as an additional radiation component $\Omega_\nu \propto a^{-4}$ early on, but during adiabatic cooling with the expansion of the Universe they become non-relativistic and the energy density behaves like ordinary matter $\Omega_\nu \propto a^{-3}$ today. The small contribution from Eq.~\ref{eq:Omega_nu} to the overall energy budget also implies that their effect on the background expansion history is small.

Their weak interaction cross-section makes neutrinos a dark matter component. However, compared to the standard cold dark matter, they have considerable bulk velocities. This changes the growth of perturbations on scales smaller than the distance travelled by neutrinos up to today, the neutrino horizon, defined by
\begin{equation}
    d_\nu(t_0) = \int_{t_\mathrm{ini}}^{t_0} c_\nu (t') \mathrm d t' \: ,
\end{equation}
with the average neutrino velocity $c_\nu$, which is close to the speed of light early on. The neutrino horizon itself is numerically closely related to the more commonly used free-streaming wavenumber at the time of the non-relativistic transition, $k_\mathrm{nr}$ \citep{Lesgourgues_neutrino_book}
\begin{equation}
\label{eq:k_nr}
k_\mathrm{nr} \approx 0.0178 \: \Omega_m^{1/2} \left(\frac{m_\nu}{\mathrm{eV}} \right)^{1/2} \: \mathrm{Mpc}^{-1} \: h \: .
\end{equation}
On scales exceeding the neutrino horizon, velocities can be neglected and the perturbations consequently evolve identical to those in the cold dark matter component. For smaller scales $k \gg k_\mathrm{nr}$ within the neutrino horizon, however, free-streaming leads to slower growth of neutrino perturbations. Due to gravitational backreaction on the other species, this causes a characteristic step-like suppression of the linear matter power spectrum approximately given by \cite{Hu_1998}
\begin{equation}
\left. \frac{P_{\nu}}{P} \right|_{k \gg k_\mathrm{nr}} \approx 1 - 8 \frac{\Omega_\nu}{\Omega_m} \: .
\end{equation}
To compare the density power spectrum between cosmologies with and without neutrinos, we here assumed the same primordial perturbations and kept the total $\Omega_m$ (including neutrinos) fixed, resulting in equal positions of the peak of the power spectrum and ensuring that the spectra are identical in the super-horizon limit. The cosmologies for our neutrino simulations described in Sec.~\ref{sec:simulations} are chosen in the same way.

The interplay between neutrinos and $f(R)$ gravity is interesting due to a curious coincidence: the typical range of the fifth force given by the Compton wavelength $\mu^{-1}$ in Eq.~\ref{eq:Poisson_large_field} and the free-streaming scale of neutrinos in Eq.~\ref{eq:k_nr} are comparable for the relevant parameter space of neutrino masses and values of $|f_{R0}|$, such that the known standard model neutrinos might counteract signatures of boosted growth caused by modified gravity. This makes neutrinos important for constraints on $f(R)$, and this paper searches for ways to disentangle both effects.

\subsection{The DUSTGRAIN-{\em pathfinder} simulations}
\label{sec:simulations}

\begin{table*}
\begin{center}
\begin{tabular}{lcccccccc}
Simulation Name & Gravity type  &  
$|f_{R0}| $ &
$\sum m_{\nu }$ [eV] &
$\Omega _{CDM}$ &
$\Omega _{\nu }$ &
$M^{p}_{CDM}$ [M$_{\odot }/h$] &
$M^{p}_{\nu }$ [M$_{\odot }/h$] & $\sigma_8$ \\
\hline \hline
$\Lambda $CDM & GR & -- & 0 & 0.31345 & 0 & $8.1\times 10^{10}$ & 0 & $0.842$\\
fR4 & $f(R)$  & $ 10^{-4}$ & 0 & 0.31345 & 0 & $8.1\times 10^{10}$  & 0 & $0.963$ \\
fR5 & $f(R)$  & $ 10^{-5}$ & 0 & 0.31345 &0  & $8.1\times 10^{10}$  & 0 & $0.898$ \\
fR6 & $f(R)$  & $ 10^{-6}$ & 0 & 0.31345 & 0 & $8.1\times 10^{10}$  & 0 & $0.856$ \\
fR4\_0.3eV & $f(R)$  & $ 10^{-4}$ & 0.3 & 0.30630 & 0.00715 & $7.92\times 10^{10}$ & $1.85\times 10^{9}$ & $0.887$ \\
fR5\_0.15eV & $f(R)$  & $ 10^{-5}$ & 0.15 & 0.30987 & 0.00358 & $8.01\times 10^{10}$ & $9.25\times 10^{8}$ & $0.859$ \\
\hline
\end{tabular}
\caption{Summary of the main numerical and cosmological parameters characterising the subset of the DUSTGRAIN-{\em pathfinder} simulations considered in this work. In the table, $M^p_{\nu}$ represents the neutrino simulation particle mass, $M^p_{CDM}$ represents the CDM simulation particle mass, while $\Omega_{CDM}$ and $\Omega_{\nu}$ the CDM and neutrino density parameters, respectively. The listed $\sigma_8$ values represent the linear power normalisation attained at $z=0$, while all simulations are normalised to the same spectral amplitude ${A}_{s}=2.199\times 10^{-9}$ at the redshift of the CMB.}
\label{tab:simulations}
\end{center}
\end{table*}

Our analysis is based on a subset of the DUSTGRAIN-{\em pathfinder} simulations suite described in \cite{Giocoli_Baldi_Moscardini_2018}. The main purpose of the DUSTGRAIN-{\em pathfinder} simulations is to explore the degeneracy between neutrino and modified gravity (MG) effects by sampling the joint $f(R)-\sum m_{\nu }$ parameter space with combined $N$-body simulations that simultaneously implement both effects in the evolution of cosmic structures. To this end, the {\small MG-GADGET} code -- specifically developed by \citet{Puchwein_Baldi_Springel_2013} for $f(R)$ gravity simulations -- has been combined with the particle-based implementation of massive neutrinos described in \citet{Viel_Haehnelt_Springel_2010}, allowing to include a separate family of neutrino particles to the source term of the $\delta f_R$ field equation \ref{eq:field_equation_fR}, which then reads:
\begin{equation}
\label{eq:field_equation_fR_nu}
\nabla^2 \delta f_R = \frac{a^2}{3} \Big( \delta R(f_R) - 8 \pi G \delta \rho_{CDM} - 8 \pi G \delta \rho_{\nu }\Big)\; .
\end{equation}

The DUSTGRAIN-{\em pathfinder} simulations follow the evolution of $(2\times )768^3$ particles of dark matter (and massive neutrinos) in a periodic cosmological box of $750\; h^{-1}$ Mpc per side from a starting redshift of $z_{i}=99$ to $z=0$, for a variety of combinations of the parameters $|f_{R0}|$ in the range $\left[ 10^{-6}, 10^{-4}\right] $ and $\sum m_{\nu }$ in the range $\left[ 0.0, 0.3\right]$ eV, plus a reference $\Lambda $CDM simulation (i.e. GR with $\sum m_{\nu }=0$). The cosmological parameters assumed in the simulations are consistent with the Planck 2015 constraints \citep[see][]{Planck_2015_XIII}: $\Omega _{M}=\Omega _{CDM}+\Omega _{b}+\Omega _{\nu} = 0.31345$, $\Omega _{\Lambda }=0.68655$, $h=0.6731$, $\sigma _{8}(\Lambda \mathrm{CDM})=0.842$. The dark matter particle mass (for the massless neutrino cases) is $M_{CDM}=8.1\times 10^{10}\; h^{-1}$ M$_{\odot }$ and the gravitational softening is set to $\epsilon _{g}= 25\; h^{-1}$kpc, corresponding to $(1/40)$ times the mean inter-particle separation.

Initial conditions for the simulations have been generated by following the Zel'dovich approximation to generate a random realisation of the linear matter power spectrum obtained with the Boltzmann code {\small CAMB}\footnote{www.cosmologist.info} \citep[][]{camb} for the cosmological parameters defined above and under the assumption of standard GR. For the simulations including massive neutrinos, besides updating the {\small CAMB} linear power spectrum used to generate the initial conditions accordingly, we also employ the approach described in \citet{Zennaro_etal_2017, Villaescusa-Navarro_etal_2018} which amounts to generating two fully correlated random realisations of the linear matter power spectrum for standard Cold Dark Matter particles and massive neutrinos based on their individual transfer functions. Neutrino thermal velocities are then randomly sampled from the corresponding Fermi distribution and added on top of gravitational velocities to the neutrino particles.
The same random seeds have been used to generate all initial conditions in order to suppress cosmic variance in the direct comparison between models. As the simulations start at $z_{i}=99$ when $f(R)$ effects are expected to be negligible, no modifications are necessary to incorporate them in the initial conditions and the standard GR particle distributions -- with and without neutrinos -- can be safely employed for both the GR and $f(R)$ runs.

A summary of the main parameters of the simulations considered in this work is presented in Table~\ref{tab:simulations}. We refer the interested reader to \cite{Giocoli_Baldi_Moscardini_2018} for a more detailed description of the DUSTGRAIN-{\em pathfinder} simulations.

\section{Cosmic Degeneracies}
\label{sec:results}

\begin{figure*}
  \includegraphics[width=1.\textwidth]{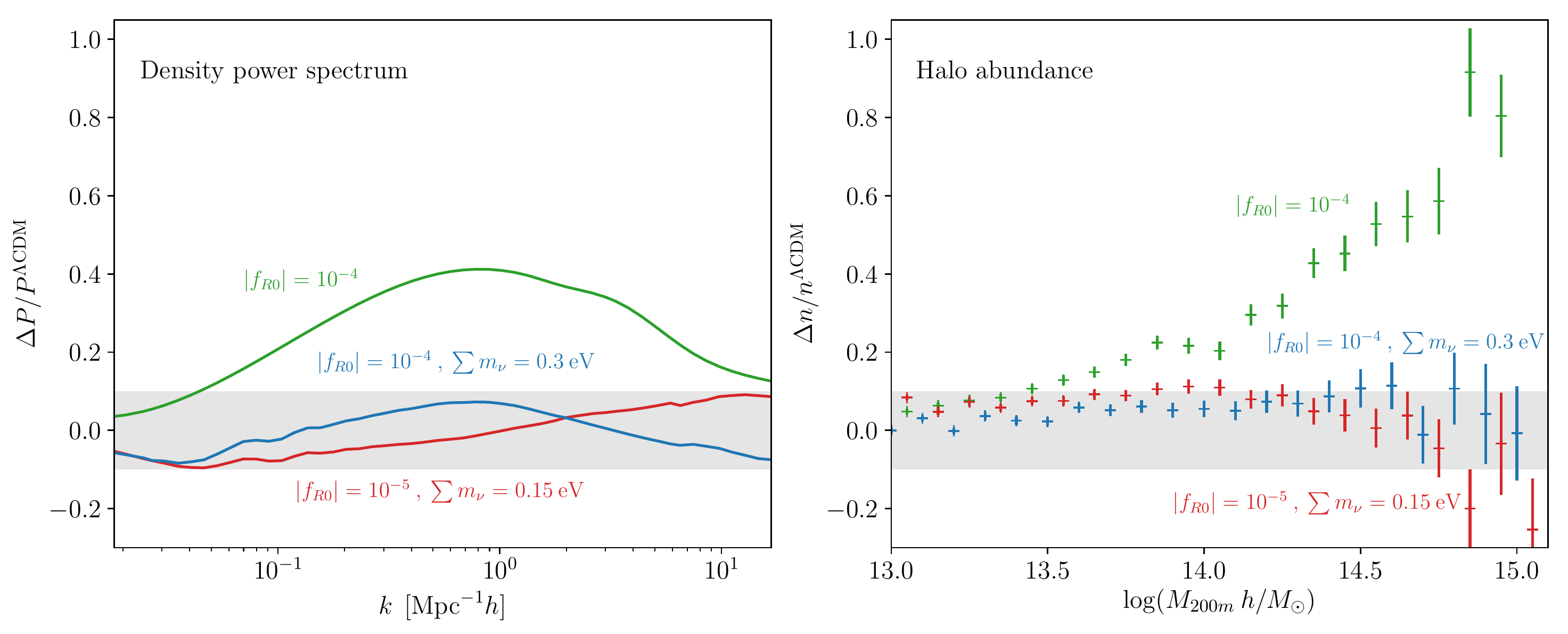}
\caption[Caption]{\textbf{Left:} Relative deviation induced by $f(R)$ gravity and massive neutrinos in the matter power spectrum measured in a subset of our simulations at $z=0$. The large deviation caused by the additional growth in $|f_{R0}|=10^{-4}$ is almost completely counteracted by massive neutrinos with $\sum m_\nu = 0.3 \: \mathrm{eV}$. We find a similar case for $|f_{R0}| = 10^{-5}$ and $\sum m_\nu = 0.15 \: \mathrm{eV}$. \textbf{Right:} The same degeneracy in the simulated abundance of halos at $z=0$. Note that the degeneracy is non-trivial, the same $P(k)$ can lead to different cluster abundances in $f(R)$ since the collapse threshold is changed in modified gravity. The uncertainty for the cluster abundance is calculated with Poisson error bars. Shaded grey bands indicate the $10 \%$ deviation region in both plots.}%
\label{fig:degeneracy}
\end{figure*}

The first $N$-body simulation to investigate the joint effects of neutrinos and modified gravity was performed in \cite{Baldi_etal_2014} where the authors pointed out the degeneracy between the competing signals. This was confirmed by multiple recent papers based on simulations to study how neutrinos can mask $f(R)$ imprints in the kinematic Sunyaev-Zeldovich effect of massive galaxy clusters \citep{Roncarelli_Baldi_Villaescusa-Navarro_2016, Roncarelli_Baldi_Villaescusa-Navarro_2018}, in weak lensing statistics \citep{Giocoli_Baldi_Moscardini_2018, Peel_etal_2018} and in the abundance of galaxy clusters \citep{Hagstotz_2018}.
A first attempt to exploit Machine Learning techniques to separate the two signals was put forward by \cite{Peel_etal_2018b, Merten_etal_2018}.

All these studies confirm a degeneracy in observables that rely on structure growth, which makes the unknown neutrino masses an important nuisance parameter when constraining $f(R)$ gravity, as pointed out in \cite{Hagstotz_2018}. These papers also show that especially the redshift evolution can be a potentially powerful tool in distinguishing these models since the time evolution of the modifications induced by $f(R)$ and neutrinos differs in general. However, many large-scale structure data sets available today do not have sufficient redshift reach to set stringent constraints on deviations from general relativity while marginalising over neutrino mass.

We refer to the above cited papers for details how these degeneracies play out for various probes and how they can be broken with higher redshift data, but the main challenge is summarized in Fig.~\ref{fig:degeneracy}, where we show the relative change induced in the matter power spectrum (left) and the halo abundance (right). Note that even though the halo mass function is clearly derived from the matter power spectrum, the degeneracy in the cluster abundance demonstrated here is non-trivial since the threshold of collapse $\delta_c$ also changes in $f(R)$ gravity \citep[e.g.][]{Schmidt2009, Kopp2013, Cataneo2016, Braun_Bates2017}. Within current observational accuracy, the effect of modified gravity leading to additional structure growth and the suppression effect of neutrino free-streaming are thus difficult to distinguish. Therefore, extending the cosmological parameter space with free neutrino masses tends to weaken existing limits on $|f_{R0}|$.

Since the degeneracy is broken by the different redshift evolution of the density $\delta$ in $f(R)$ and neutrino cosmologies, it is interesting to consider the growth rate of structures to tell them apart. In linear theory, the continuity equation
\begin{equation}
    \frac{\partial \delta}{\partial t} + \frac{1}{a} \nabla \cdot \mathbf{v} = 0 \: ,
\end{equation}
relates the growth rate $f = \mathrm d \ln D_+ / \mathrm d \ln a$ directly to the velocity divergence
\begin{equation}
\theta = \frac{1}{H} \nabla \cdot \mathbf v = - a \delta f \: ,
\end{equation}
which we use as a probe of the different growth histories in GR, modified gravity and neutrino cosmologies. We then investigate the degeneracy between the latter in two regimes:
\begin{itemize}
\item The large-scale velocity divergence 2-point function in Fourier space $P_{\theta \theta}$ as a proxy for the growth rate. We present the detailed results in Sec.~\ref{sec:2-point}.
\item The velocity dispersion inside of non-linear collapsed structures in Sec.~\ref{sec:clusters}
\end{itemize}

\subsection{Velocity divergence 2-point functions}
\label{sec:2-point}

We compute the velocity dispersion $\theta = 1/H \: \nabla\cdot\mathbf{v}$ and interpolate it on a uniform, $512^3$-point grid, using the publicly available \texttt{DTFE} code \citep{Cautun_2011}. 

\begin{figure}
	\includegraphics[width=\columnwidth]{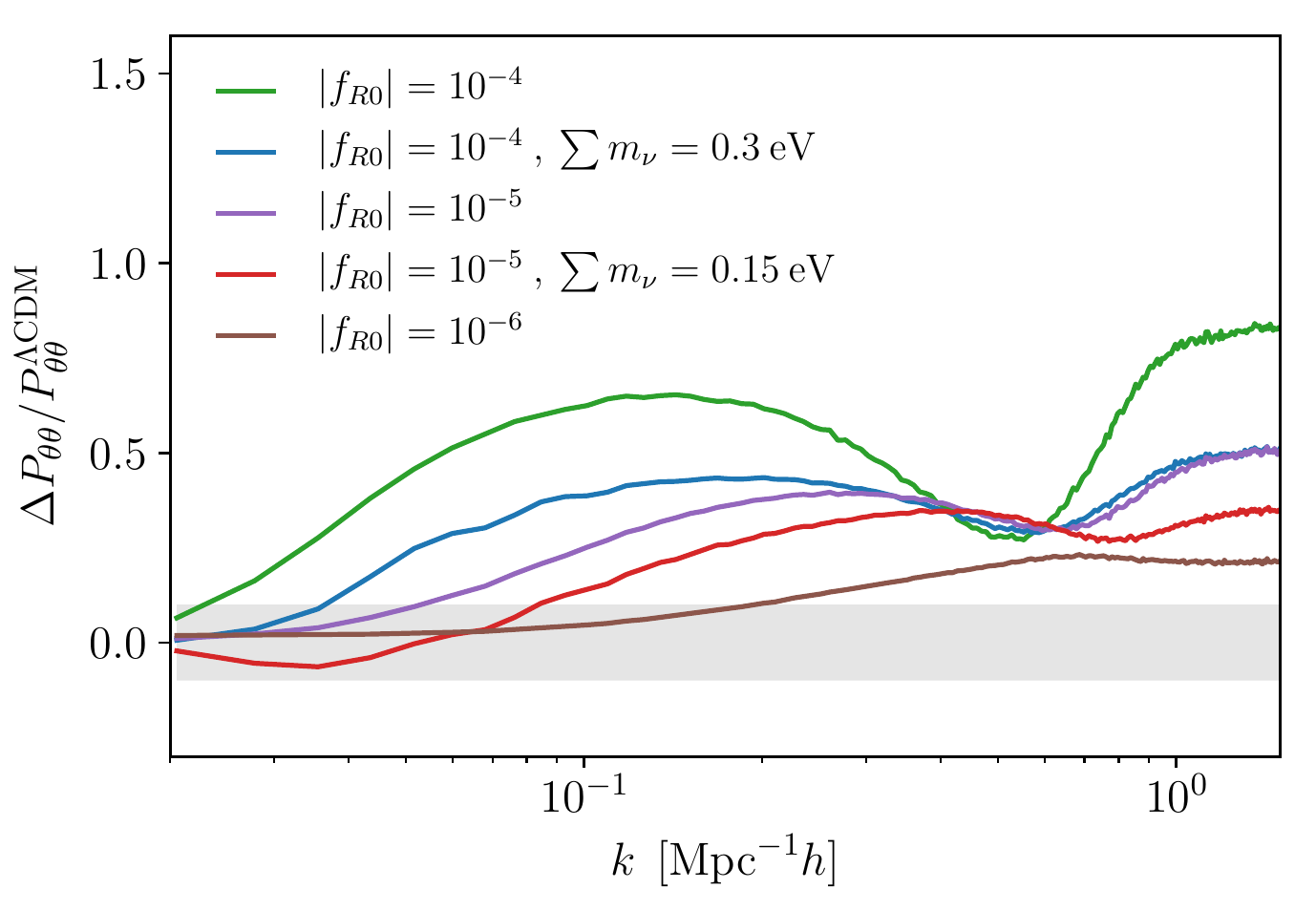}
    \caption{Relative change in the velocity divergence power spectrum $P_{\theta \theta}$ compared to $\Lambda$CDM for various models with modified gravity, massive neutrinos, or both. The deviation from $\Lambda$CDM is more pronounced compared to the approximately degenerate density power spectra for combinations of $|f_{R0}|$ and $\sum m_\nu$ shown in Fig.~\ref{fig:degeneracy}. The dip in the spectra marks the onset of collapsed structures. The shaded band indicates a $10 \%$ deviation range.}
    \label{fig:velocity_divergence_pk}
\end{figure}

This allows us to compare the power spectrum $P_{\theta \theta}$ in the $\Lambda$CDM simulation with the $f(R)$ and massive neutrino simulations in Fig.~\ref{fig:velocity_divergence_pk} where we plot (as in the left panel of Fig.~\ref{fig:degeneracy} for the matter power spectrum) the relative deviation from the $\Lambda$CDM value. 
Clearly, all modified gravity simulations show an increased velocity divergence -- and therefore growth rate -- on scales $\gtrsim 0.1\,\mathrm{Mpc}\,h^{-1}$, with the $|f_{R0}|=10^{-4}$ simulation showing the strongest enhancement since the fifth force becomes active first. Very large scales $k \ll \mu^{-1}$ exceeding the range of the force given by the Compton wavelength of the scalar field are not affected. These results confirm previous findings \citep[see e.g.][]{Jennings_etal_2012} that the velocity power spectrum provides a much stronger signature of modified gravity compared to the density power spectrum, thereby representing a more powerful tool to test gravity on cosmological scales. In principle it can be probed by redshift space distortion measurements sensitive to $f \sigma_8 / b$ with the tracer bias $b$ \citep{Peacock_2001, BOSS_DR12_2017}. However, the scale dependence of $f$ in modified gravity, changes in galaxy formation and subsequently the tracer bias and difficult modelling of the nonlinear effects in modified gravity make this analysis challenging \citep[see the discussion in][]{Jennings_etal_2012, Hernandez_2018}.

The addition of neutrinos (cf. the two $|f_{R0}|=10^{-5}$ runs in Fig.~\ref{fig:degeneracy}) dampens the velocity divergence field slightly overall, but unlike for the density power spectrum this effect is not sufficient to counteract the enhanced growth rate in $f(R)$. This confirms the redshift evolution of the degeneracy in the density field: at early times $z \gtrsim 0.5$, $f(R)$ effects are small, and neutrino suppression of the matter fluctuations dominates. As soon as the additional force enhancement becomes active, it tends to win out and we arrive at the approximate degeneracy observed in Fig.~\ref{fig:degeneracy} today. In the future evolution, $f(R)$ effects will dominate over the neutrino damping for the cases shown here.

The plot also demonstrates that hierarchical formation of collapsed objects in $f(R)$ proceeds faster than in a $\Lambda$CDM universe. Small structures form first, and this process proceeds to larger scales with time. Since the fifth force accelerates the collapse, cosmologies with higher values of $|f_{R0}|$ contain larger nonlinear structures at a given redshift $z$. The transition to these collapsed structures appears as a characteristic dip in the velocity divergence power spectrum \citep[see also the detailed explanation in][]{Li_etal_2013a}.

\subsection{Cluster velocity dispersion}
\label{sec:clusters}

We now turn to the kinematics inside of non-linear structures. The velocity dispersion of galaxy cluster members is a long-established measure of the total gravitational potential via the virial theorem, and therefore it can serve as a mass proxy of the system \citep{Biviano_2006}. First studies of $f(R)$ effects on virialised systems were presented by \cite{Lombriser_2012_virial}, and recently efforts have been made to use the phase space dynamics of single massive clusters to constrain modified gravity \citep[e.g.][]{Pizzuti2017}.

Here we focus on the change in the mean observable velocity dispersion instead of detailed studies of single objects. Starting point is the virial theorem, which itself is a consequence of phase-space conservation expressed by the Liouville equation and holds for any system obeying Hamiltonian dynamics. It is therefore unchanged by $f(R)$ gravity, and states in its scalar form
\begin{equation}
    2 E_\mathrm{kin} + E_\mathrm{pot} = 0 \: ,
\end{equation}
with kinetic and potential energy of the system respectively. From there, we can get a rough estimate for the velocity dispersion 
\begin{equation}
\label{eq:sigma_sq_}
    \sigma^2 \approx \frac{G M(r)}{r}
\end{equation}
for a virialised system of size $r$. This makes the velocity dispersion a direct measurement of the gravitational potential of a bound system. For an unscreened cluster in $f(R)$, Eq.~\ref{eq:Poisson_large_field} leads to an enhancement of the gravitational force and potential by a factor $4/3$ -- we therefore expect the velocity dispersion to be boosted by $(4/3)^{1/2}$ compared to the standard prediction.

\begin{figure}
	\includegraphics[width=\columnwidth]{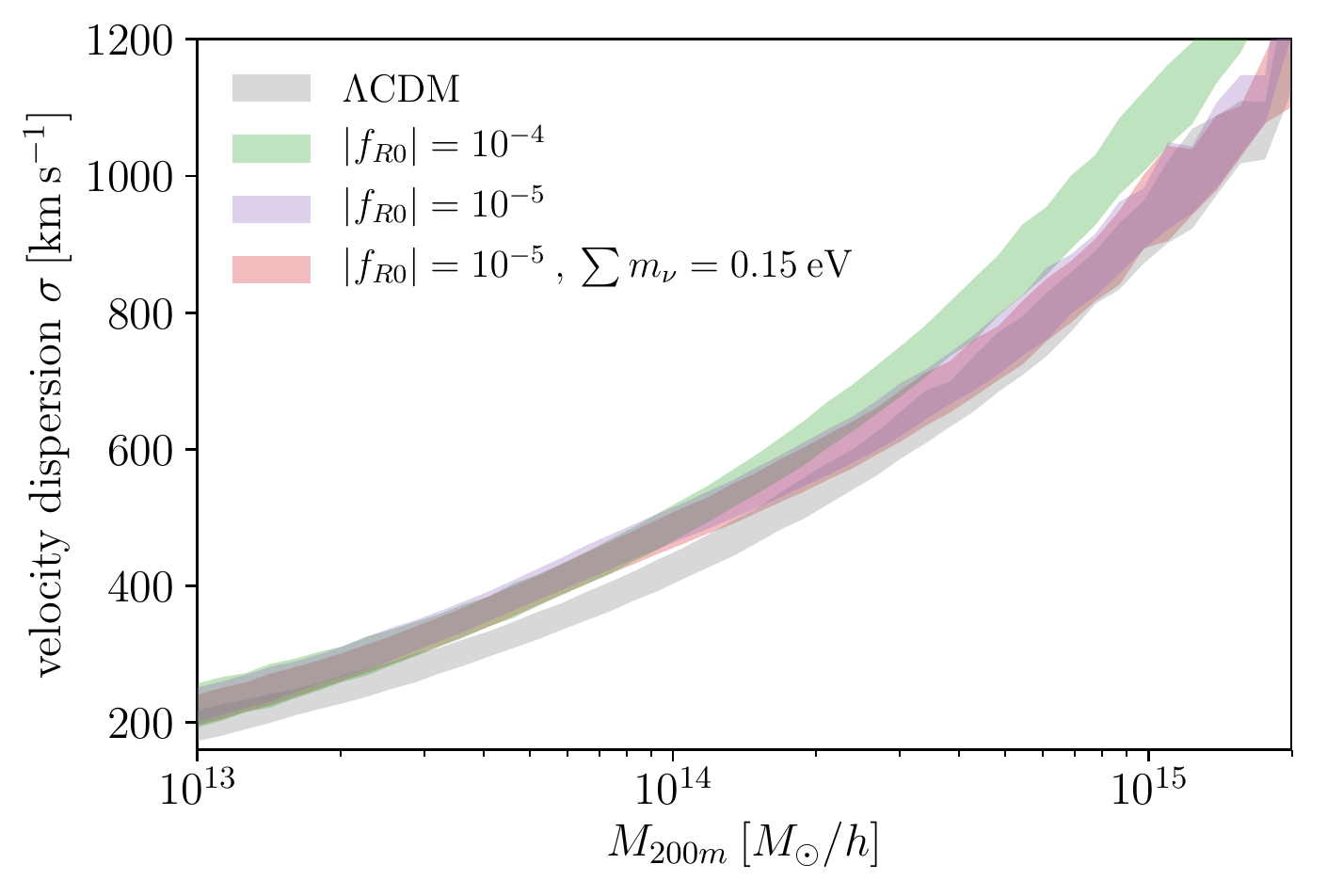}
    \caption{Velocity dispersion $\sigma$ within clusters of a given mass $M_{200m}$ for a subset of the studied cosmologies at $z=0$. Shaded region shows the standard deviation found in our simulations. Note that most systems are virialised, either to the $\Lambda$CDM value or the boosted unscreened $f(R)$ equilibrium. Neutrinos do not have any detectable effect on the velocity dispersion inside of clusters, and we just show the case with $|f_{R0}| = 10^{-5}$ and $\sum m_\nu = 0.15~\mathrm{eV}$ for clarity. The relative deviations are shown separately in Fig.~\ref{fig:rel_velocity_dispersion_sim}.}
    \label{fig:velocity_dispersion}
\end{figure}

However, the screening mechanism of $f(R)$ gravity outlined in Sec.~\ref{sec:fR} is crucial to understand the full phenomenology of the theory. We can estimate the mass scale of objects with potential wells deep enough to activate the screening mechanism with the condition set by Eq.~\ref{eq:thin_shell}. In order to do that, we consider the force enhancement caused by $f(R)$
\begin{equation}
    g(r) \equiv \frac{\mathrm d \psi / \mathrm d r}{\mathrm d \psi_N / \mathrm d r}
\end{equation}
relative to the Newtonian potential $\psi_N$. We can from there calculate the average \textit{additional} potential energy of the system
\begin{equation}
\label{eq:force_enhancement}
    \bar g = \frac{\int \mathrm d r r^2 w(r) g(r)}{\int \mathrm d r r^2 w(r)} \: ,
\end{equation}
which varies between 1 (for the screened case) and $4/3$ (for the unscreened case), with the weighting function
\begin{equation}
    w(r) = \rho(r) r \frac{\mathrm d \psi_N}{\mathrm d r} \: .
\end{equation}
Following \cite{Schmidt_2010}, we assume that the additional force is only sourced by the mass distribution beyond the \textit{screening radius} $r_\mathrm{screen}$, which is defined by the equality in condition Eq.~\ref{eq:thin_shell}, i.e. 
\begin{equation}
\label{eq:r_screen}
    \frac{2}{3} \psi_N(r_\mathrm{screen}) = \bar f_{R}(z) \: .
\end{equation} 
This implies for the force enhancement
\begin{equation}
    g(r) = 1 + \frac{1}{3} \frac{M(<r) - M(<r_\mathrm{screen})}{M(<r)} \: ,
\end{equation}
and by assuming NFW density profiles we can solve the equations above to determine $\bar g$. We use the concentration-mass relation by \cite{Bullock2001} to fix the density profiles, but the overall results for $\bar g$ are rather insensitive to the specific choice of $c(M, z)$. From the modified potential energy, the virial theorem then suggests the scaling of the velocity dispersion $\sigma$ in $f(R)$ as
\begin{equation}
    \frac{\sigma^{f(R)}}{\sigma^{\Lambda \mathrm{CDM}}} \propto \bar g^{1/2} \: .
\end{equation}
The screening radius $r_\mathrm{screen}$ itself depends on time via the evolution of the density profile $c(M, z)$ and the background evolution of the scalar field
\begin{equation}
    \bar f_R(z) = | f_{R0} | \frac{1 + 4\frac{\Omega_\Lambda}{\Omega_m}}{(1+z)^3 + 4 \frac{\Omega_\Lambda}{\Omega_m}} \: .
\end{equation}

The velocity dispersion measured in our simulations at $z=0$ is plotted in Fig.~\ref{fig:velocity_dispersion}, where the width of the contours represents the standard deviation found among the objects. Most of the clusters virialise either to the $\Lambda$CDM equilibrium or the boosted $f(R)$ value, and since the maximum force enhancement is identical for all models, $|f_{R0}|$ merely determines at which mass scale the transition between the two cases occurs. We also show results for the simulation with $|f_{R0}| = 10^{-5}$ and $\sum m_\nu = 0.15~\mathrm{eV}$ as an example of a cosmology with both modified gravity and massive neutrinos, but note that neutrinos have no detectable effect on the cluster velocity dispersion. Therefore the dynamics of galaxies within clusters are an excellent way to break the degeneracy found in measurements relying on the amplitude of the matter fluctuations.

\begin{figure}
	\includegraphics[width=\columnwidth]{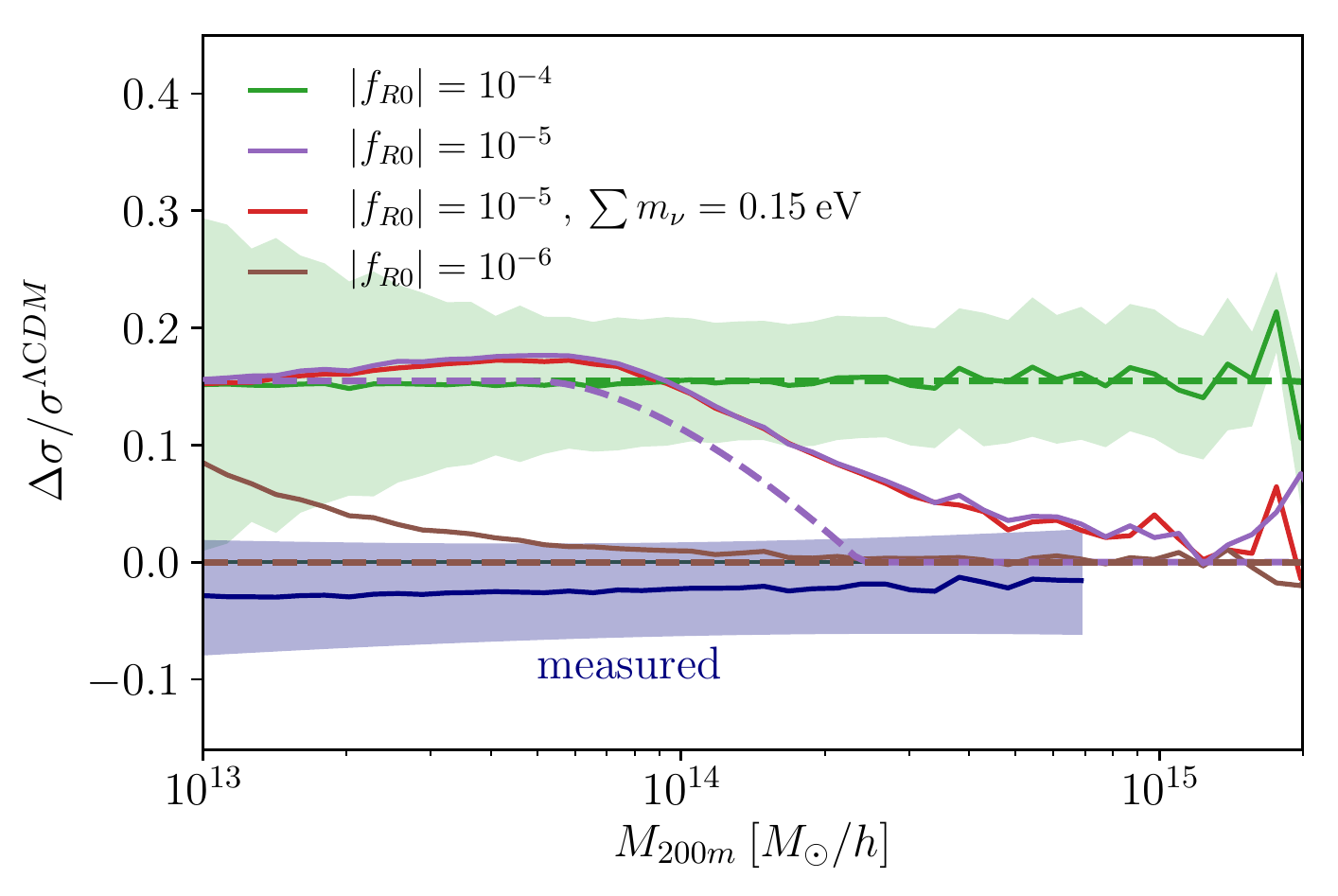}
    \caption{Relative velocity dispersion within clusters of a given mass in the extended cosmologies, normalised to the mean value of the $\Lambda$CDM simulation. The (propagated) error bar of the ratio $\Delta \sigma / \sigma$ is showcased for the $|f_{R0}| = 10^{-4}$ model as shaded region, and has a similar magnitude for all curves. The other error bars are suppressed for clarity. Also shown is the empirical relation (blue) with propagated error bars as described in the text. Dashed lines show the expectation $\Delta \sigma / \sigma \approx \bar g^{1/2}$ from the simplified force enhancement model in Eq.~\ref{eq:force_enhancement}. For unscreened clusters, the velocity dispersion is larger by a factor $\sqrt{4/3} \approx 1.15$ as expected from the virial theorem in $f(R)$.}
    \label{fig:rel_velocity_dispersion_sim}
\end{figure}

We focus on the relative deviations from $\Lambda$CDM in Fig.~\ref{fig:rel_velocity_dispersion_sim}, where we normalise the curves to the values measured in our fiducial simulation. Dashed lines show the prediction $\Delta \sigma / \sigma \approx \bar g^{1/2}$ from Eq.~\ref{eq:force_enhancement}.

Clusters for $|f_{R0}| = 10^{-4}$ are all unscreened, and virialise to the $f(R)$ equilibrium value boosted by a factor $(4/3)^{1/2} \approx 1.15$. On the other hand $|f_{R0}| = 10^{-6}$ is almost completely screened, and just shows slight deviations for low mass systems with $M_{200m} \sim 10^{13} M_\odot h^{-1}$. The intermediate case $|f_{R0}| = 10^{-5}$ demonstrates how the screening mechanism becomes active for clusters with $M_{200m} \sim 2 \times 10^{14} M_\odot h^{-1}$ with a long transition tail towards the fully screened regime. This also implies that single very massive clusters are not well suited to constrain $f(R)$ models \citep[see e.g.][for a case study]{Pizzuti2017}.

The simple model from Eq.~\ref{eq:force_enhancement} somewhat overestimates the efficiency of the screening mechanism, in agreement with findings by \cite{Schmidt_2010}. It therefore only serves as a conservative estimate for the transition region. In addition, even clusters that are screened today can still carry the imprint of the fifth force if parts of the progenitor structures were unscreened in their past. The relaxation time of a galaxy cluster of richness $N$ is approximately given by \citep{Binney_Tremaine_book}
\begin{equation}
    t_r \approx \frac{0.1 N}{\ln N} t_\mathrm{cross}
\end{equation}
with typical crossing times $t_\mathrm{cross}\approx 1~\mathrm{Gyr}$, this leads to relaxation timescales of order $t_r \approx 2~\mathrm{Gyr}$ for a richness $N \sim 100$ and can range up to the Hubble time $t_r \approx 14.5~\mathrm{Gyr}$ for very massive clusters with $N\sim 1000$ member galaxies.

We also compare the results found in the simulations to an empirical $\sigma(M)$ relation which we obtained by combining the mass-richness relation of \cite{Johnston_2007} and the $\sigma$-richness relation of  \cite{Becker_2007}. Both studies used the catalog of the Sloan Digital Sky survey \citep[SDSS;][]{2009ApJ...703.2217S} which allowed us to combine the two empirical relations.
The uncertainty shown in Fig.~\ref{fig:rel_velocity_dispersion_sim} is the (propagated) uncertainty quoted in \cite{Johnston_2007} and \cite{Becker_2007}.

Even without giving a quantitative upper limit on $f_{R0}$ here, we note that the $|f_{R0}| = 10^{-5}$ results seem to be incompatible with the observed cluster velocity dispersion irrespective of neutrino effects. This is comparable to current upper limits obtained from large-scale structure data \citep[e.g.][]{Cataneo2014}.


\section{Conclusions}
\label{sec:conclucsion}


Neutrinos are of great interest for modified gravity searches in the large-scale structure since they suppress the growth of structures on scales comparable to the range of the fifth force expected in deviations from GR. The uncertainty in the neutrino mass scale leads to an uncertainty in the size of this suppression, which can mask the characteristic additional growth of structures in $f(R)$ gravity. This degeneracy was studied before in the context of the amplitude of matter fluctuations and found to be time dependant, since the modifications in the growth of structures induced by neutrinos and the fifth force have different redshift dependencies.

Therefore, in this paper we studied the velocity divergence power spectrum $P_{\theta \theta}$ in Sec.~\ref{sec:2-point} as a proxy for the linear growth rate. Compared to $\Lambda$CDM it is strictly enhanced in our simulations at $z=0$, also in cosmologies including both modified gravity and massive neutrinos that show a comparable amplitude of matter fluctuations at that time. We conclude that for combinations of parameters that show approximate degeneracy in the matter power spectrum today, neutrino suppression dominates in the past, while in the future evolution the additional growth induced by the fifth force will win out. This effect can be probed by redshift-space distortion measurements, but an analysis accounting for the scale dependant growth in $f(R)$ remains challenging \citep{Jennings_etal_2012, Hernandez_2018}.

As a second step, we studied the kinematics inside of clusters in Sec.~\ref{sec:clusters}. The velocity dispersion found in our simulations agrees well with the expectations from the virial theorem, and it is enhanced in the unscreened $f(R)$ regime by a factor $(4/3)^{1/2}$ proportional to the the maximum force enhancement. Neutrinos on the other hand do not have any detectable effect on the velocity dispersion. Since the free-streaming length is larger than the typical cluster size, they behave as a smooth background component. So while they suppress the overall cluster abundance, the kinematics inside of halos are completely unaffected. We also compare the simulated dynamics to the empirical $\sigma - M$ relation found by combining the results from \cite{Johnston_2007} and \cite{Becker_2007} and find good agreement with the $\Lambda$CDM simulation. While we do not quote a stringent upper limit on the modified gravitiy parameter $|f_{R0}|$, we point out that the observed relation is in strong tension with expectations from an $|f_{R0}| = 10^{-5}$ model for clusters of mass $M_{200m} \approx 10^{-14} M_\odot h^{-1}$ -- independent  of the  neutrino mass.

Overall, kinematic information is an excellent observable to detect fifth force effects irrespective of the unknown neutrino mass. 
Using  kinematic information could also be potentially useful in order to break other degeneracies with (screened) modified gravity theories such as baryonic feedback processes stemming, e.g., from AGNs  which also reduce clustering \citep{Arnold_Puchwein_Springel_2014,2018A&A...615A.134E}.

\begin{acknowledgements}
Many cosmological quantities in this paper were calculated using the Einstein-Boltzmann code \texttt{CLASS} \citep{CLASS}.

We appreciate the help of Ben Moster with cross-checks for our simulation suite and helpful discussions with Raffaella Capasso on cluster dynamics. SH acknowledges the support of the DFG Cluster of Excellence ”Origin and Structure of the Universe” and the Transregio programme TR33 ”The Dark Universe”. 
MG was supported by by NASA through the NASA Hubble
Fellowship grant \#HST-HF2-51409 awarded by the Space Telescope Science
Institute, which is operated by the Association of Universities for
Research in Astronomy, Inc., for NASA, under contract NAS5-26555.
MB acknowledges support from the Italian Ministry for Education, University and Research (MIUR) through the SIR individual grant SIMCODE (project number RBSI14P4IH), from the grant MIUR PRIN 2015 ”Cosmology and Fundamental Physics: illuminating the Dark Universe with Euclid”, and from the agreement ASI n.I/023/12/0 “Attivita` relative alla fase B2/C per la missione Euclid”. The DUSTGRAIN-pathfinder simulations discussed in this work have been performed and analysed on the Marconi supercomputing machine at Cineca thanks to the PRACE project SIMCODE1 (grant nr. 2016153604, P.I. M. Baldi) and on the computing facilities of the Computational Centre for Particle and Astrophysics (C2PAP) and the Leibniz Supercomputing Centre (LRZ) under the project ID pr94ji.

We thank the Research Council of Norway for their support. Some computations were performed on resources provided by UNINETT Sigma2 -- the National Infrastructure for High Performance Computing and Data Storage in Norway. This paper is partly based upon work from the COST action CA15117 (CANTATA), supported by COST (European Cooperation in Science and Technology).
\end{acknowledgements}

\bibliographystyle{aa}
\bibliography{Bibliography}

\end{document}